\documentclass[a4paper,fleqn,usenatbib,twocolumn]{mnras}

\usepackage{newtxtext,newtxmath}

\usepackage[T1]{fontenc}
\usepackage{ae,aecompl}

\usepackage{graphicx}	
\usepackage{amsmath}	
\usepackage{amssymb}	
\usepackage[usenames,dvipsnames]{color} 

\title[The CMB and the IMF]{The Cosmic Microwave Background and the Stellar Initial Mass Function}

\author[Adam S. Jermyn et al.]{
Adam S. Jermyn,$^{1}$\thanks{E-mail: adamjermyn@gmail.com}
Charles L. Steinhardt$^{2}$
and Christopher A. Tout$^{1}$
\\
$^{1}$Institute of Astronomy, University of Cambridge, Madingley Rd, Cambridge CB3 0HA, UK\\
$^{2}$Cosmic Dawn Center, Niels Bohr Institute, Blegdamsvej 17, 2100 K\o benhavn, Denmark
}

\date{Accepted XXX. Received YYY; in original form ZZZ}

\pubyear{2017}

\begin{document}
\label{firstpage}
\pagerange{\pageref{firstpage}--\pageref{lastpage}}
\maketitle

\begin{abstract}

We argue that an increased temperature in star-forming clouds alters the stellar initial mass function to be more bottom-light than in the Milky Way.
At redshifts $z \gtrsim 6$, heating from the cosmic microwave background radiation produces this effect in all galaxies, and it is also present at lower redshifts in galaxies with very high star formation rates (SFRs).
A failure to account for it means that at present, photometric template fitting likely overestimates stellar masses and star formation rates for the highest-redshift and highest-SFR galaxies.
In addition this may resolve several outstanding problems in the chemical evolution of galactic halos. 

 \end{abstract}




\section{Introduction}

Recent ultradeep surveys~\citep{CANDELS, Steinhardt2014a,Bouwens2015,Bouwens2016, Laigle2016} have measured the rest-frame ultraviolet (UV) luminosity functions for the most luminous galaxies to redshifts of $6 < z < 10$.  These studies find a substantial population of UV-bright galaxies at high-redshift, so that a survey should expect to find several galaxies per $100~\mathrm{arcmin}^2$ at $z = 8$ and even one $z \approx 10$ galaxy brighter than 26th (AB) magnitude in the H band (rest-frame UV for $8< z < 10$ galaxies).  This population presents a rich target environment for followup observations over the next few years on the James Webb Space Telescope~\citep{JWST}.

However, connecting these measurements with theory is a far more difficult proposition.  Theoretical models of galaxy assembly predominantly describe the dark matter halo \citep{Press1974,Sheth2001} rather than the baryons which provide the measured luminosity.  Further, even nascent attempts to include baryons in halo simulations \citep{Somerville2015,Illustris} describe galaxies in terms of stellar mass ($M_\mathrm{H}$) and star formation rates, which require several additional assumptions to convert to UV luminosity ($L_\mathrm{UV}$).  These include a stellar initial mass function (IMF), dust abundance, composition and corresponding extinction law and even a star formation history.  These are difficult to constrain even for local galaxies and, as a result, there is substantial uncertainty in the $M_*/L_\mathrm{UV}$ and $M_\mathrm{H}/L_\mathrm{UV}$ ratios at high redshift.

Improving our understanding of these processes at high redshift has recently become critical because of the remarkable abundance of UV-bright galaxies at high redshift. Use of abundance matching~\citep[cf.][]{Behroozi2015} to find a correspondence between the halo mass function produced by the standard cosmological $\Lambda$ Cold Dark Matter ($\Lambda$CDM) paradigm and observed luminosity functions, $M_\mathrm{H}/L_\mathrm{UV}$ must decrease sharply for $z > 6$ from the $0 < z < 4$ ratio.  It has been proposed that this might be due to an increased stellar baryon fraction~\citep{Finkelstein2015}, increased star formation efficiency~\citep{Trac2015} or additional extinction~\citep{Mashian2016} at high redshift, each of which would change $M_*/L_\mathrm{UV}$ and therefore the inferred $M_\mathrm{H}/L_\mathrm{UV}$ as well.  However, there are currently no theoretical models to explain why these mass-to-light ratios should decrease sharply at $z \approx 6$ after remaining nearly constant at lower redshifts.  If $M_\mathrm{H}/L_\mathrm{UV}$ at $z > 6$ is the same as at $z = 4$, the existence of these luminous and hence massive early galaxies would be strongly inconsistent with $\Lambda$CDM~\citep{Steinhardt2016}.  

A top-heavy (or bottom-light) IMF at $z > 6$ could also change the stellar mass-to-light ratio. However, local dwarf galaxies with less than $1\,$percent solar metallicity, lower than expected for $z \approx 6$ galaxies, have an IMF consistent with the Milky Way \citep{Fagotto1994,Dias2010}.  Thus, metallicity-driven changes in the IMF are likely reserved for redshifts well above $z=6$.

However, metallicity is not the only relevant variable.
Processes which increase the temperature of star-forming molecular clouds could also alter the IMF~\citep{1976MNRAS.176..367L,1998MNRAS.301..569L} and hence the stellar mass-to-light ratio.
We show that if the IMF depends on temperature then cosmic microwave background (CMB)-driven and cosmic ray-driven heating of these clouds should alter the IMF for all galaxies, independently of properties or environment, at $z \gtrsim 6$ as well as for the galaxies with the highest star formation rates (SFRs) at lower redshifts.
This is consistent with the findings of~\citet{2001MNRAS.324..484H}, who inferred the variation of the IMF from Milky Way halo stars, and with theoretical arguments by~\citet{2003MNRAS.343.1224C}, who investigated this in relation star formation and halo collapse.
Our focus here is specifically on the implications for observations of high-redshift galaxies, particularly with regards to the mass-to-light ratio, though we also explore other tests of these effects.

Unfortunately the small-scale physical processes which affect star formation are still not well understood and so models of the initial mass function remain either empirical or phenomenological~\citep[see e.g.][]{2014prpl.conf...53O,doi:10.1146/annurev-astro-082708-101642, Oey2011}.
Nevertheless, such models are sufficient to capture the underlying physics and scaling laws and so have proven quite useful.
Along these lines, in section~\ref{sec:sf}, we develop a temperature-dependent, bottom-light IMF along with a discussion of the physics of molecular cloud fragmentation.
This has been previously explored in the context of the Jeans mass~\citep{Desika1} with encouraging results, though we find that a somewhat different scaling is more likely relevant.
We emphasise that this is just one such model, and discuss alternative scaling laws in the same section.
All of these laws are compatible with our later analysis, but we focus on the fragmentation mass scaling in the rest of this work because that is one of the better-understood models.

In section~\ref{sec:pop_cmb} we examine the implications of this IMF for stellar populations and the mass-to-light ratio of early galaxies, accounting for CMB heating.
In section~\ref{sec:pop_cr} we attempt similar modelling for cosmic ray (CR) heating.
This model is incomplete because it neglects feedback between the shape of the IMF and the SFR--gas temperature relationship but it suffices to highlight the expected magnitude of the CR effect.
We then discuss various observational tests of this model in section~\ref{sec:obs} as well as existing evidence and conclude with a discussion of possible complications in section~\ref{sec:comp}.

We emphasise that our intention is to provide a simple model of these phenomena to emphasise the potential importance and effects of CMB and CR heating, particularly at high redshift and in extreme environments.
This we hope will motivate more detailed studies of these phenomena.

\section{Temperature Dependence of Star Formation}
\label{sec:sf}

Although a full treatment of star formation is very complex and would require modelling many different baryonic processes~\citep[cf.][]{1985MNRAS.214..379L}, the observed qualitative and quantitative features of the initial mass function can be reproduced with a much simpler model~\citep{2007prpl.conf..149B}.
\begin{itemize}
\item The gas in molecular clouds is characterised by its temperature and density.  These two quantities define a mass above which the cloud is unstable to gravitational collapse, known as the Jeans mass~\citep{Jeans1}.  Star formation begins when a cloud exceeds its Jeans mass and begins to collapse.
\item In the early stages the cloud is optically thin and efficiently cools.  As a result, it remains isothermal through this collapse, with its temperature set by that of the ambient radiation field.  It is straightforward to show that this means the Jeans mass decreases as the cloud becomes denser and so the collapse continues unimpeded~\citep{1985MNRAS.214..379L}.
\item However, at some point, the cloud becomes optically thick.  When this occurs the cloud collapses adiabatically rather than isothermally, causing the Jeans mass to rise to meet the cloud mass~\citep{1976MNRAS.176..367L}, halting the collapse. This effect is crucial because an isothermal collapse is never halted by gas pressure~\citep[see e.g.][]{2018A&A...611A..89L}.
\item An ultimate cutoff on the final fragment mass distribution is set by the minimum mass $\tilde{m}$, known as the minimum fragmentation mass, at which a cloud can cool efficiently~\citep{1976MNRAS.176..367L}.
\end{itemize}

This description is sufficient to match both observations and simulations well \citep{2007prpl.conf..149B}.
Of particular note are that there is a knee in the IMF at the Jeans mass at which clouds are forced to be isothermal\ \citep{2006MNRAS.368.1296B}, initial separations of single stars are of the order of the Jeans length\ \citep{2002ApJ...578..914H} and few stars are seen below the minimum fragmentation mass\ \citep{2005MNRAS.356.1201B}.
In addition, the IMF power law above the knee is consistent with scale-free gravitational collapse and fragmentation\ \citep{2001ApJ...556..837K} and below the knee it is consistent with the results of simulations\ \citep{2005MNRAS.356.1201B}.

Our goal in this work is to determine how this process should be affected by changing the cloud temperature $T$, so that we can determine the effect of CMB-driven heating at $z \gtrsim 6$, where the CMB temperature exceeds the $20\,\mathrm{K}$ or so in typical star-forming molecular clouds in the Milky Way \citep{0004-637X-684-2-1228}.  To that end, a key prediction of this model is that the characteristic mass scale of the IMF is set not by the initial cloud mass but rather by this minimum fragmentation mass.  That is, the IMF ought to depend not on the mass $m$ of a star but rather on the dimensionless quantity $m/\tilde{m}$.  
Therefore we expect that the higher-temperature IMF $\xi(m,T)$ behaves as a rescaled function of mass, such that
\begin{align}
	\frac{dN}{dm}\left(m,T\right) = \xi(m,T) = g\left(\frac{m}{\tilde{m}(T)}\right),
\end{align}
where $N$ is the number of stars, or equivalently
\begin{align}
\xi(m,T) = g\left(\frac{m}{f(T)\tilde{m}(T_0)}\right),
\label{eq:scale}
\end{align}
where
\begin{align}
f(T) \equiv \frac{\tilde{m}(T)}{\tilde{m}(T_0)}
\end{align}
is the temperature rescaling function and $T_0$ is a reference temperature.

In practice equation~\eqref{eq:scale} gives a formula for turning an observed initial mass function into one with a different ambient temperature, as long as the temperature giving rise to the observed function is also known.
Given $\tilde{m}(T)$ and $T_0$ we may pick our favourite observational IMF, rescale the mass according to $\tilde{m}(T)$ and obtain a new IMF, which we expect to be valid at a different temperature.
In this work we choose the IMF of~\citet{2001MNRAS.322..231K} because it is integrable and has a readily-interpreted pair of kinks close together and near the minimum fragmentation mass in the Milky Way.
This IMF is
\begin{align}
\xi(m) = \frac{dN}{dm}\propto \begin{cases}
m^{-0.3}, & m < a_1 \tilde{m} \\
m^{-1.3}, & a_1 \tilde{m} < m < a_2 \tilde{m}\\
m^{-2.3}, & a_2 \tilde{m} < m,
\end{cases}
\label{eq:newIMF}
\end{align}
where $a_1$ and $a_2$ are dimensionless constants.  
These are assumed to be universal such that the minimum fragmentation mass is the only relevant mass scale.
Matching equation\ \eqref{eq:newIMF} to the $z=0$ IMF of~\citet{2001MNRAS.322..231K} we find that
 \begin{align}
	a_2 = \frac{0.5}{0.08} a_1 = 6.25 a_1
\end{align}
and
\begin{align}
	a_1 \tilde{m}_0 = 0.08 M_\odot,
\end{align}
where $\tilde{m}_0$ is the present-day Milky Way minimum fragmentation mass.
Note that for masses below approximately $0.08M_\odot$ the objects are not stars in the sense of fusing but we include them in our calculations because they still contribute to the condensed baryonic mass of a stellar population.

When the ambient temperature is large enough it serves to regulate the cooling of collapsing clouds.
Because cooling is the limiting factor for fragmentation this ultimately regulates the fragmentation mass, such that 
\begin{align}
	 \tilde{m} = \frac{\kappa}{2}\left(\frac{\pi k_\mathrm{B} T}{G \mu}\right)^2 = 1.5\times 10^{-3} M_\odot \left(\frac{\mu}{m_{\rm p}}\right)^{-2}\left(\frac{\kappa}{\kappa_0}\right) \left(\frac{T}{\rm K}\right)^2
     \label{eq:mt}
\end{align}
\citep{1976MNRAS.176..367L}, where $k_\mathrm{B}$ is the Boltzmann constant, $G$ is the gravitational constant, $\mu$ is the mean molecular weight, $m_{\rm p}$ is the proton mass, and $\kappa$ is the opacity of the cloud and $\kappa_0$ is the opacity of ionized hydrogen.
This is the case for
\begin{align}
	T > T_\mathrm{c} \approx 4.1 \left(\frac{\kappa_0}{\kappa}\right)^{4/7} \mathrm{K}.
	\label{eq:condition}
\end{align}
Here we take $\kappa = \kappa_0$ but $T_{\rm c}$ is sufficiently smaller than the temperatures of interest, which generally exceed $20\,\mathrm{K}$, that even if this were not the case then equation~\eqref{eq:condition} would still be satisfied.
In this case the rescaling function takes on the simple form
\begin{align}
f(T) = \left(\frac{T}{T_0}\right)^2.
\end{align}
This is in agreement with the simulations performed by~\citet{2009MNRAS.392..590B}.

Putting this all together, the initial mass function becomes
\begin{align}
\xi(m,T) = \frac{dN}{dm}(T)\propto \begin{cases}
m^{-0.3}, & m < 0.08 M_\odot f(T) \\
m^{-1.3}, & 0.08 M_\odot f(T) < m < 0.50 M_\odot f(T)\\
m^{-2.3}, & 0.50 M_\odot f(T) < m,
\end{cases}
\label{eq:newIMFfull}
\end{align}
where the proportionality constants are such that $\xi(m,T)$ is continuous.

A variety of other models have been proposed for the dependence of the mass scale on temperature.
\citet{2005A&A...435..611J} found in simulations that $f(T) \propto T^{3/2}$.
\citet{2005MNRAS.356.1201B} likewise propose $f(T) \propto T^{3/2}$ but with an additional dependence on density.
\citet{2012MNRAS.423.2037H} proposes $f(T) \propto T$ with an additional dependence on the sonic radius as defined therein.
Yet another scaling relation is provided by~\citet{2011ApJ...743..110K}, who suggests that $f(T) \propto T^{-1/18}$ with an additional dependence on density.
Neglecting the variation of the density and sonic radius each of these relations fits well into our formalism and we are agnostic as to which ought to be preferred\footnote{We may expand $f(T)$ to include these other parameters but such an analysis is left for the future.}.
For simplicity we proceed with the model described by equation~\eqref{eq:newIMFfull} but note that our analysis is straightforwardly extended to other models.

\section{CMB Heating}
\label{sec:pop_cmb}

\begin{figure}
	\centering
	\includegraphics[width=0.45\textwidth]{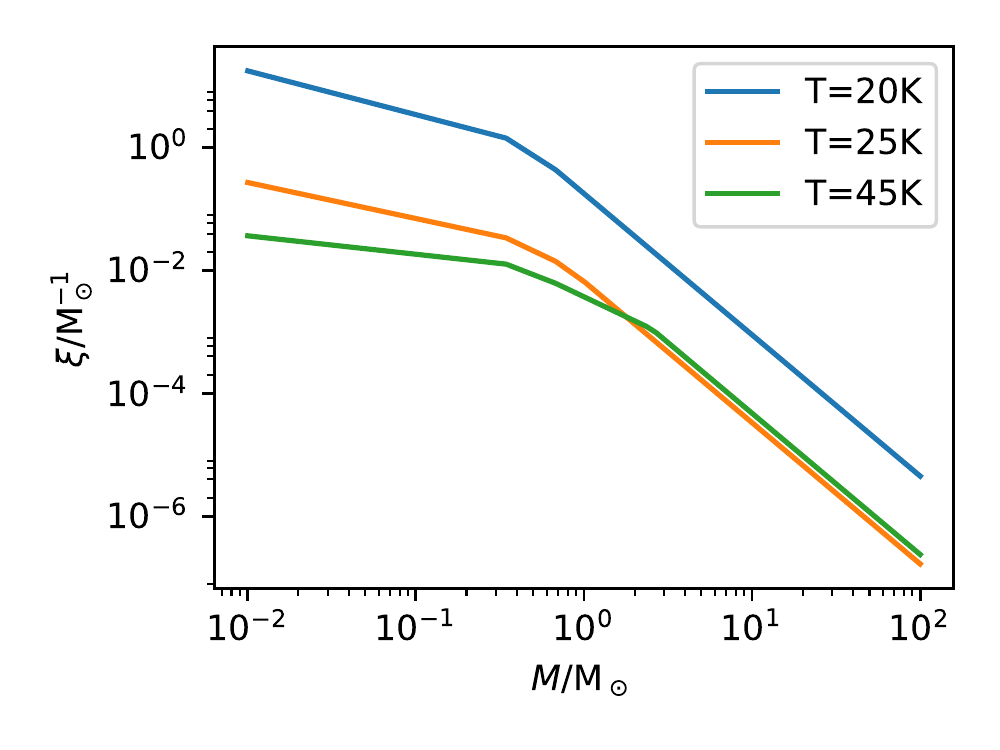}
	\caption{The IMF described by equation~\eqref{eq:newIMFfull} is shown for several temperatures, each with the same total mass, corresponding to the CMB at $z=6.3$ ($20\,$K), 8.3 ($25\,$K), and 15.5 ($45\,$K). As $z$ increases, the breaks in the IMF power-law shift to higher masses.  Other phenomena that increase the minimum gas temperature in star-forming regions will have the same effect.}
	\label{fig:imf}
\end{figure}

The dependence of $\tilde{m}$ on the ambient temperature $T$ leads to the remarkable conclusion that, even at modest redshift, the CMB temperature $T_{\mathrm{CMB}}$ becomes relevant~\citep{2005MNRAS.359..211L,0004-637X-715-1-194} and gives rise to scaling with redshift $z$ of the form
\begin{align}
	\tilde{m} \propto T^2 \propto (1+z)^2
\end{align}
for
\begin{align}
	z > z_c = \frac{T_\mathrm{cloud}}{T_{\mathrm{CMB,0}}} - 1 \approx 6.3,
\end{align}
where $T_\mathrm{CMB,0}$ is the current temperature of the CMB and $T_\mathrm{cloud}$ is the background temperature the molecular cloud would otherwise have, here taken to be $T_\mathrm{cloud} = \mathrm{20\,K}$.
This means that, if all other physics remains the same, we should expect all mass scales to shift upward with redshift, with the possible exception of the cloud mass, which is determined by the large-scale dynamics and contents of the galaxy rather than by smaller-scale thermodynamics.
So 
\begin{align}
 f(z) \equiv f(T(z)) = \min\left(1, \frac{1+z}{6.3}\right)^2.
\end{align}
Thus, the redshift dependence of the IMF can be expressed solely in terms of the redshift dependence of $\tilde{m}$. 
As $z$ increases, the breaks in the power-law IMF shift towards higher masses (Fig. \ref{fig:imf}).

\subsection{Effects on Inferred Quantities}
\label{sec:pop_cmb}

If the IMF is indeed bottom-light at high gas temperatures compared with the local Universe, every quantity currently inferred for these galaxies (which include all $z \gtrsim 6$ galaxies) with the assumption of a static IMF have been incorrectly estimated.  Here, we attempt to estimate the magnitude of the possible corrections to key quantities used to describe the first galaxies.

It would be ideal to simply perform a new analysis of high-redshift photometric catalogues with spectra generated from a variable IMF.  The potential effects on inferred quantities could then be calculated directly from a comparison with the previous catalogue.  However, current photometric template-fitting codes are not designed to support this sort of link between the the IMF and redshift or to track the history of the stellar population in a way that allows for the effects of a time-dependent IMF.  So the full calculation is not straightforward.  Instead we can get a good idea of the effects by calculating the stellar population and its effect on the two most widely used inferred quantities (apart from redshift) for high-redshift galaxies, stellar mass $M_*$ and star formation rate (SFR).

First the IMF (equation~\ref{eq:newIMFfull}) must be turned into a stellar population. 
Two additional ingredients, namely the history of the star formation rate $\dot{m}_\mathrm{SFR}(t)$ as well as the lifetime of stars $\tau_\mathrm{s}(m)$, are needed to do this.
For the SFR we use the prescription
\begin{equation}
	\log_{10} \frac{\dot{M}_{\mathrm{SFR}}}{M_\odot \mathrm{GYr}^{-1}} = \left(0.84-0.026\frac{t}{\mathrm{GYr}}\right)\log_{10} \frac{M_*}{M_\odot} - 6.51 + 0.11\frac{t}{\mathrm{GYr}}
\end{equation}
\citep{0004-637X-796-1-25}.
This gives a mass-independent SFR at early times, approximately exponential growth later on and finally quiescence at low redshift.

Next we obtain $\tau_{\rm s}$ from scaling relations.
With stellar main--sequence luminosities of~\citep{1992isa..book.....B}
\begin{align}
	L_\mathrm{s}(m) \approx L_\odot \left(\frac{m}{M_\odot}\right)^{3.5} 
\end{align}
and lifetime energy released $E \propto m$~\citep{1989ApJ...347..998E}, we find an effective lifetime of~\citep{1992isa..book.....B}
\begin{align}
	\tau_{\mathrm{s}}(m) \approx 10^{10} \left(\frac{m}{M_\odot}\right)^{-2.5}\textrm{ yr}.
\end{align}
This is not quite correct owing to deviations in both the luminosity and lifetime for high stellar masses~\citep{1989ApJ...347..998E,1996MNRAS.281..257T}, but it does a good job of reflecting the fact that the specific energy released over the lifetime of a star is approximately independent of mass.

To put the pieces together we integrate star formation over time.
To first order, stellar remnants may be neglected when producing light curves, so that the relevant stellar population has a mass profile approximated by
\begin{align}
	\eta(m,t) = \frac{dN}{dm} &= \int_0^t \xi(m, z(t'))\dot{M}_{\mathrm{SFR}} (t') H(\tau_\mathrm{s}(m) + t' - t) dt'\\
	&= \int_{\max(0,t - \tau_\mathrm{s}(m))}^t \xi(m, z(t'))\dot{M}_{\mathrm{SFR}} (t') dt',
	\label{eq:pop}
\end{align}
where $H$ is the Heaviside step function.

\begin{figure}
	\includegraphics[width=0.5\textwidth]{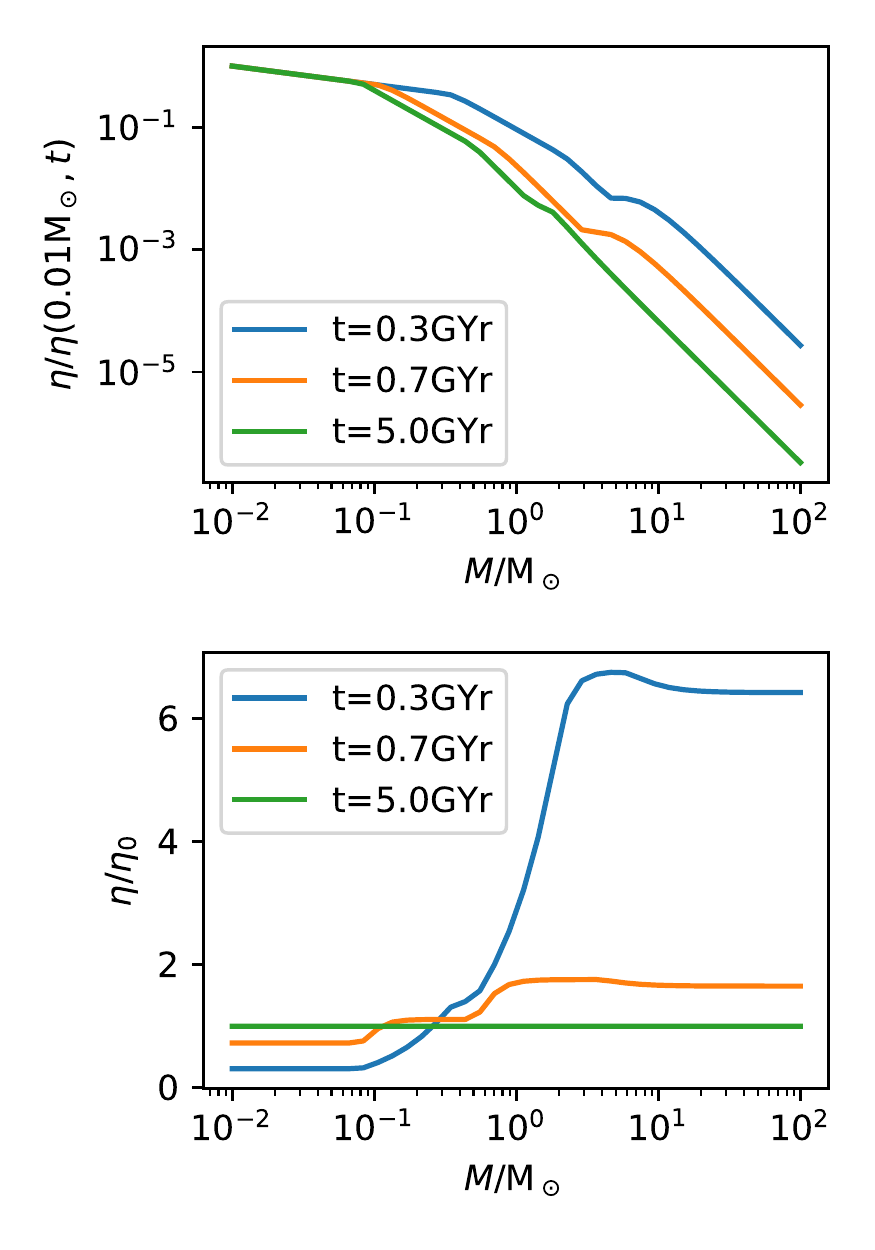}
	\caption{(a) A sample stellar population mass distribution is shown for several times, normalised in each case to the low end of the mass range. At early times the IMF is more top-heavy owing to suppression of the low end of the mass range.  (b) The same distributions, normalized to the population $\eta_0$ that would result from a static IMF.}
	\label{fig:dist}
\end{figure}

The stellar mass distribution $\eta(m,t)$ becomes significantly top-heavier (or bottom-lighter) within the first Gyr after the Big Bang (Fig.~\ref{fig:dist}) because the CMB suppresses the low end of the mass range.
The excess is sharpest at higher masses, because that is where the IMF slope is most extreme, but drops off at the mass corresponding to when the stellar lifetime equals the elapsed time since initiation of star formation in the galaxy.  Thus, the peak in the excess moves to lower masses as time goes on.

Although a full treatment of the effect on the inferred stellar mass requires a modified photometric template fitting code, it can be estimated by examination of the mass-to-light ratio.
As with the static IMF, stellar populations produced by the IMF in equation~\eqref{eq:newIMFfull} have a luminosity~\citep{1992isa..book.....B}
\begin{align}
	L = \int_{M_{\rm min}}^{M_{\rm max}} \eta(m,t) m^{3.5} dm,
\end{align}
where $M_{\rm min} = 0.08 M_\odot$ is the minimum stellar mass and $M_{\rm max}$ is the maximum stellar mass.
Importantly, this luminosity is dominated by the upper end of the mass distribution.
In contrast, the mass of this population
\begin{align}
	M_* = \int_{M_{\rm min}}^{M_{\rm max}} \eta(m,t) m dm,
\end{align}
is dominated by the cutoff $0.08M_\odot f(z)$ at which the exponent of the mass crosses $-1$.
In total, the effects of the CMB produce a top-heavier IMF and thus stellar population than a static $z=0$ IMF and so produce a higher luminosity for the same amount of mass.  Failure to account for this effect results in an overestimate of the mass-to-light ratio and so an overestimate of both the masses and star formation rates of high-redshift galaxies.

\begin{figure}
	\includegraphics[width=0.45\textwidth]{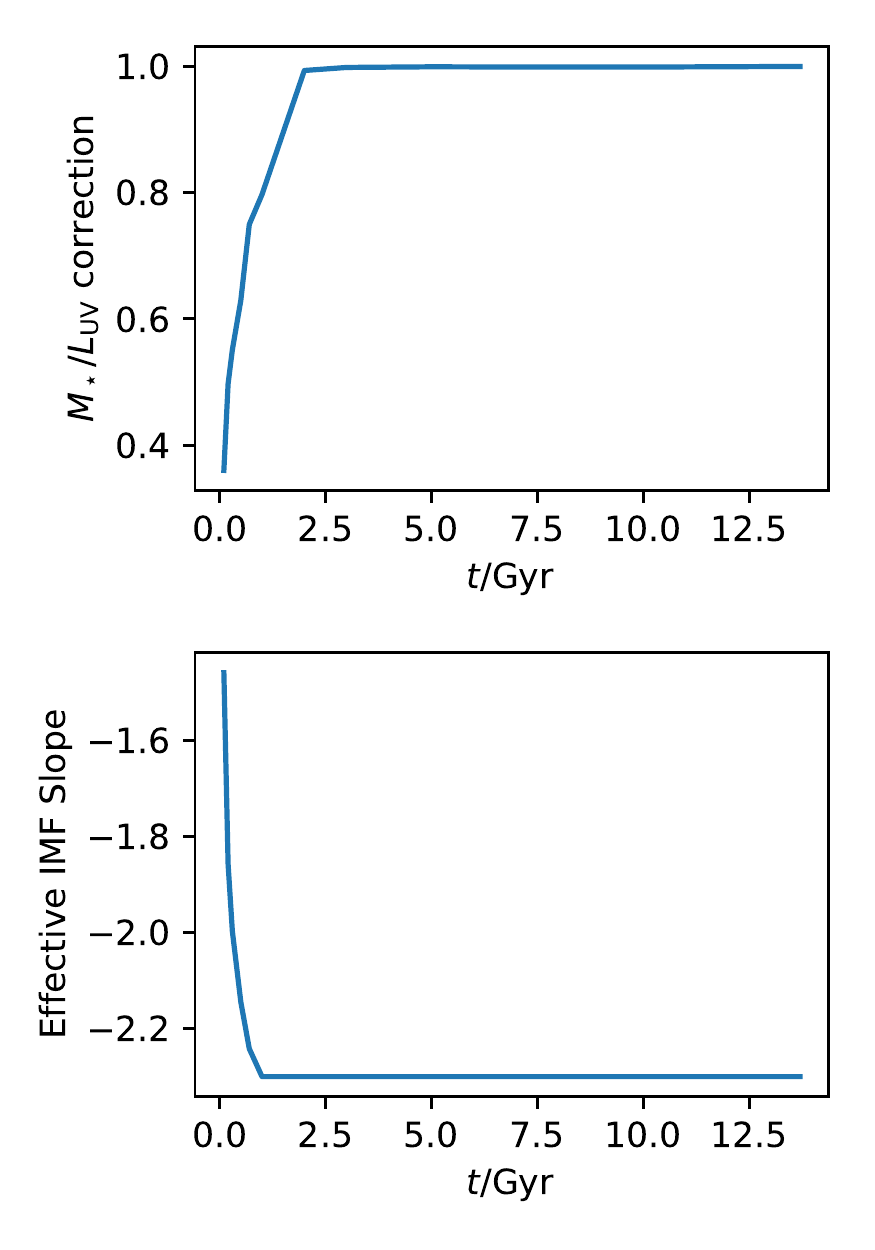}
	\caption{(top) The ratio of the mass-to-light ratio for the time-dependent IMF to that for the time-independent IMF is shown as a function of time for our fiducial cosmology at 3000\,\AA~(UV, proxy for star formation rate). (bottom) The effective IMF slope between the Milky Way knee of $0.5 M_\odot$ and $50 M_\odot$ is shown as a function of time for the same cosmology. Note that at high redshift the discrepancy in mass-to-light ratio and effective slope becomes quite large and is in the direction needed to resolve the impossibly early galaxy problem.  Both would be overestimated at $z \gtrsim 6$ if the effects of CMB temperature are neglected.}
	\label{fig:ml1}
\end{figure}

To estimate this we calculate the correction to the stellar mass-to-light ratio for monochromatic luminosity at rest-frame wavelengths of 3000\,\AA~as a function of time for our fiducial cosmological model, shown in the top panel of Fig.~\ref{fig:ml1}\footnote{The correction is nearly independent of wavelength. Although the mass-to-light ratio varies sharply with wavelength, the flux at all wavelengths is dominated by high-mass stars and to leading order the correction just tracks the change in this population.}.
The bottom panel of the same shows the effective slope of the time-dependent IMF computed between the Milky Way knee of $0.5 M_\odot$ and $50 M_\odot$.
The stellar mass-to-light ratio sharply rises in the first $\mathrm{GYr}$ or so as the effective IMF slope changes rapidly and the initial stellar population is established, so that stellar masses may be significantly overestimated during this epoch.  
This is precisely the region in which the inferred masses of high-redshift galaxies apparently require either rapid shifts in the stellar baryon fraction \citep{Finkelstein2015} or may even be impossible to produce with the $\Lambda$CDM halo mass function \citep{Steinhardt2016}.  The predicted shift in $M_*/L_\mathrm{UV}$ is in the correct direction to reduce the tension between theory and observation but this effect alone is insufficient to solve the problem entirely and an additional effect is required.

\section{Cosmic Ray Heating}
\label{sec:pop_cr}

Although the CMB provides a universal contribution to all galaxies independent of environment or stage of evolution, local effects can further increase the gas temperature in the star-forming regions of individual galaxies.  Likely the strongest effect in high-redshift star-forming galaxies comes from cosmic rays \citep{PPP2010}.  Even at $z \leq 2$, where the CMB contribution is negligible, dust temperatures in massive star-forming galaxies are observed to range from $25$ to $45$\,K \citep{Magnelli2014, 0004-637X-835-2-213}. Because dust, and therefore very likely gas~\citep{doi:10.1093/mnras/stw3270}, temperatures are found to increase towards higher specific SFRs, this is particularly relevant in starburst galaxies, where there is some evidence of a top-heavy IMF~\citep{Doane1993, 2041-8205-840-2-L11}.

\begin{figure}
	\centering
	\includegraphics[width=0.45\textwidth]{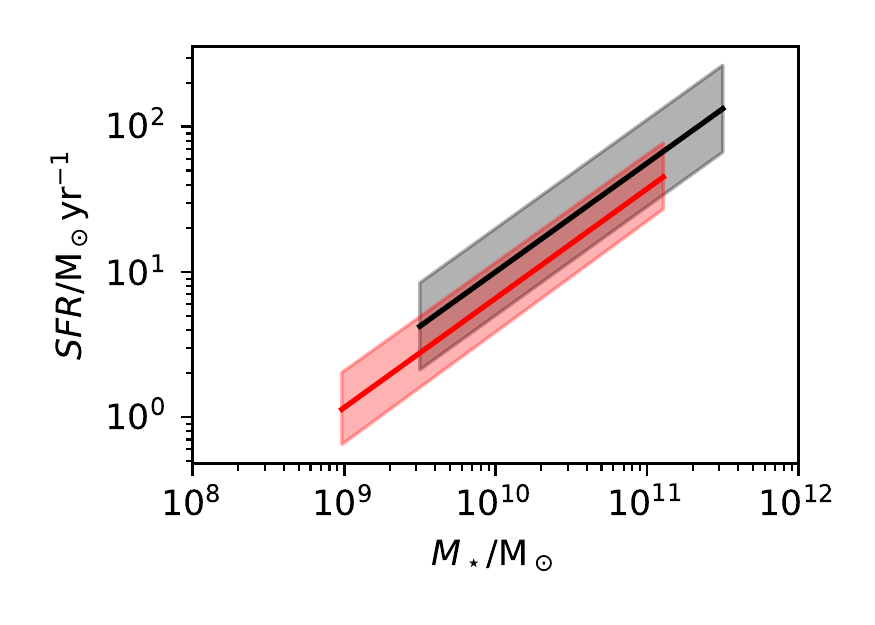}
	\label{fig:sfms}
	\caption{An idealised set of probability contours for the star-forming main sequence are shown for the currently inferred population (black) and the IMF-corrected inference (red). The centre line in each case is the median and the shaded region is the $\pm 1 \sigma$ contour. After correcting for the variability of the IMF with star formation rate the main sequence shifts both down and to the left, doing so more strongly the more massive and more rapidly star-forming the galaxy is. This means that the already narrow main sequence is actually narrower than previously inferred.}
\end{figure}

For instance if the gas temperatures in the star-forming regions of high-redshift galaxies are well-approximated by their observed dust temperatures \citep{Magnelli2014}, at $25$ to $45\,$K both the stellar mass and star formation rates are overestimated with a static IMF.  Correcting both with the temperature-dependent mass-to-light ratio derived in section~\ref{sec:pop_cmb}, we find that the star-forming main sequence may be even narrower than originally believed (Fig.~\ref{fig:sfms}).

\section{Observational Tests}
\label{sec:obs}

We predict that the IMF, and hence the stellar population, should contain a significantly larger fraction of massive stars at high redshift, or when other conditions drive an increased gas temperature, than would be predicted by a low-redshift Milky Way IMF.  It is impossible to directly measure the stellar mass distribution of high-redshift galaxies so a direct test is impossible.  However, we have described many possible indirect tests for the bottom-light IMF.
The tests with the strongest observational constraints come from well-measured, nearby galaxies.  The spectra of these galaxies exhibit discrepancies between the predicted and inferred mass-to-light ratio \citep{cite-key}, providing indirect evidence for a variable IMF~\citep{Desika1,Desika2}.

We also expect several chemical signatures because of this phenomenon.  Increasing the typical stellar mass also increases the number of massive stars that explode as supernovae at the expense of low-mass asymptotic giants (AGB stars).  The net result of this is likely to be an increase in $^{16}$O and other heavier $\alpha$--process isotopes at the expense of carbon in the combined stellar chemical yield\ \citep{1995ApJS...98..617T, 2014PASA...31...30K}.  Similarly, the increase in higher-mass AGB stars relative to low-mass stars would mean more suffer hot bottom burning during third dredge up with the consequence of increasing the $^{14}$N yield at the expense of $^{12}$C\ \citep{1995ApJS...98..617T, 2014PASA...31...30K}.  Chemical evidence of this sort has been seen in the halo of the Milky Way~\citep{2001MNRAS.324..484H,2005ApJ...625..833L, 2012A&A...547A..76P} and it is possible that more detailed studies of this and the halos of galaxies at higher redshift may further constrain the IMF.

Additional indirect evidence comes from observations of the gamma ray burst (GRB) population.  Because GRBs are thought to originate only from higher-mass stars, they serve as a probe of that population and hence as a proxy for the IMF.  Several observations suggest an evolving luminosity function~\citep{2012ApJ...754...46T,2016ApJ...825..135M}.  Unfortunately the effect of the IMF on GRB observations is largely degenerate with the effect of metallicity and these observations are currently explained by fitting the metallicity evolution~\citep{2016ApJ...817....8P}.  Nevertheless, modern cosmological simulations are capable of testing this evolution and there has been recent interest in understanding the effect of the IMF in these simulations~\citep{Guszejnov17}, so these observations may prove useful in the near future.  

It may also be possible to test our predictions with observations of galactic clusters.
Indeed~\citet{Guszejnov17} have already performed such a test by simulating the formation of a galaxy similar to the Milky Way with a variety of different IMF temperature dependences.
They find that our model with $f(T) \propto T^2$ produces a factor of several more variation in the IMF of galactic clusters than is observed, which suggests that the effect of temperature is smaller than what we have suggested.
This is certainly possible.
There are well-motivated models which propose significantly weaker dependences which could well be correct.
In which case the effects of the CMB and cosmic rays should be correspondingly smaller\footnote{In particular~\citet{Guszejnov17} find that the protostellar heating model with $f(T)\propto T^{-1/18}$ produces less than the observed variation while the other models we have discussed produce somewhat more than is observed. This favours the former but does mean that nearly all of the observed variation in the knee of the IMF among different clusters must be explained as a result of observational uncertainties rather than intrinsic scatter. It could also be that there are as-yet unknown sources of variation or that the variation other models predict is suppressed by selection effects acting in the environments conducive to star formation.}.
It is also possible that the molecular clouds which form stars do so in more uniform environments than observations of present-day clouds suggests.
For instance there is some evidence that star formation in at least some clouds requires a gravitational trigger~\citep{2015EAS....75...43L}, which could serve to reduce the scatter in their initial temperatures.
Finally it is also possible that the details of the simulation matter, particularly with regards to the criteria for star formation\footnote{This point is discussed in some detail by~\citet{Guszejnov17} along with other caveats in their section 3.1.} and the statistics of the star particles, which in this case were comparable to or larger than the clusters from which the variation has been deduced~\citep[see e.g.][for details of the observed clusters]{doi:10.1146/annurev-astro-082708-101642}.

Finally, the most direct test would be to compare the spectra of high-redshift galaxies with those predicted by the IMF (equation~\ref{eq:newIMFfull}).  To perform this test at the redshifts of interest is beyond current observational capabilities but should be possible in the near future with the James Webb Space Telescope (JWST).
The Near-Infrared Spectrograph (NIRSPEC) should enable detailed galactic spectroscopy at $z > z_\mathrm{crit} \approx 6$ and thereby measure high-redshift stellar populations~\citep{JWST, 2017arXiv170400753V}.  

As an immediate step, in principle it should be possible to fit existing photometry of high-redshift galaxies and determine whether this time-variable IMF produces a superior fit.  However, we cannot use models with a time-varying IMF in current codes without generating a new set of templates as well as implementing the requirement that the redshift of the object match the template redshift rather than that being a fully independent parameter~\citep[see e.g.][]{2009ApJ...699..486C,2010ApJ...712..833C}.  A significant restructuring of template fitting codes would be required.  One advantage in our analysis is that the expectation of minimal dust at high redshifts substantially mitigates the degeneracy between reddening and the age of the stellar population that exists at lower redshifts.

\begin{figure}
	\includegraphics[width=0.45\textwidth]{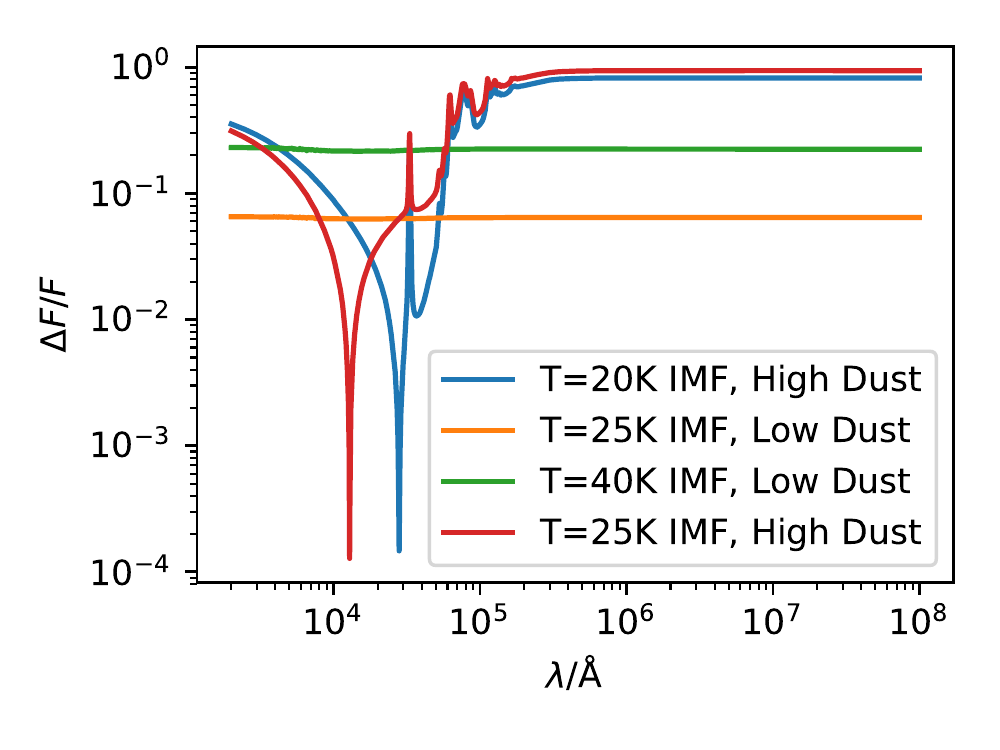}
	\label{fig:spectra}
	\caption{
	The relative flux difference $\Delta F/F$ between the standard Kroupa 2001 IMF and several modified IMFs is shown as a function of wavelength $\lambda$ for a population of age $1\,\mathrm{Gyr}$. All differences were computed between rest-frame spectra. The reference spectrum was computed with low dust content while the modified ones were computed with two different dust models (low and high). The effect of the modified IMF strongly depends on temperature and largely amounts to a rescaling. This makes it distinguishable from that of dust content in the rest-frame near infrared.}
\end{figure}

Although existing codes do not support the required templates, it is still useful to examine the spectra they predict for a single stellar population.  Simulated spectra (Fig.~\ref{fig:spectra}) were produced for a single stellar population with an age of $10^{9}\,\mathrm{yr}$ with the IMF of equation~\eqref{eq:newIMFfull} for a variety of redshifts and dust content, with the python-fsps code~\citep{2009ApJ...699..486C,2010ApJ...712..833C}.  This is what would be seen if a galaxy formed with a single starburst at the specified redshift and were then observed $10^9\,\mathrm{yr}$ later.  The effect of the modified IMF strongly depends on redshift and is distinguishable from that of dust content in the (rest-frame) near infrared but may be masked by the effects of dust at longer wavelengths.  As a result photometry including mid-IR observations should contain enough information to test these predictions if template-fitting codes can be appropriately modified.

In summary, it should be possible to conclusively determine whether this modified IMF is responsible for the spectra of high-redshift galaxies that will be observed by JWST/NIRSPEC.    Substantial modifications are required to investigate this with existing codes but, once performed, photometry of high-redshift galaxies will likely be sufficient for this purpose, particularly at high enough redshift that dust-driven extinction is minimal.

\section{Complications}
\label{sec:comp}

There are three potentially significant complications to this picture but we conclude that none is likely to fundamentally alter it.
First, if the typical age of stars increases at higher redshift then, even though the IMF would be bottom-heavy, after convolving it with the star formation history of a galaxy the net result could be to leave the mass-to-light ratio unchanged.  However, this is counter to what is both expected theoretically and observed out to $z \approx 6$ \citep{0004-637X-796-1-25} and so, if anything, likely operates in the opposite direction.

Secondly, if the typical density of molecular clouds at high redshift were once greater than today that could counteract the increase in the Jeans mass.
This is unlikely for the same reason that a decreasing cloud size is unlikely: the early molecular clouds were likely larger and more diffuse than those of our galaxy simply because there was less time available for them to collapse and develop density gradients.
However other parameters including the cloud velocity dispersion and metallicity likely also change with time and could conceivably counteract the effects of temperature variation.
Without a much more detailed analysis we cannot eliminate this possibility and so it is important to bear in mind.

Finally while we have characterised the physics of star formation by the parameters $\kappa$, $\rho$ and $T$, in practice these quantities are really drawn from a joint probability distribution.
That is, each galaxy has many molecular clouds with a variety of masses, temperatures, densities and so on, so it is an approximation to replace these distributions by typical quantities as we have done.
At $z < z_c$ this implies that there should be clouds which would have $T < T_\mathrm{CMB}(z)$ were it not for the CMB heating them, meaning that the effect ought to be observable at lower redshift than our treatment would suggest.
Likewise at $z > z_c$ there ought to be some clouds which would be hotter than $T_\mathrm{CMB}(z)$ even without CMB heating and which would therefore be unaffected either way.
This means that the effect is not quite as strong at moderate $z > z_c$ as we would predict.
The practical effect of us considering distributions rather than single representative quantities is one of smoothing the $z$ dependence of the mass-to-light ratio.
The width of this smoothing is
\begin{align}
	\Delta z \approx z_c \frac{\Delta T}{T_{\mathrm{ISM}}} = \frac{\Delta T}{T_{\mathrm{CMB}}},
\end{align}
where $\Delta T \approx 11\,\mathrm{K}$ is the spread in molecular cloud temperatures\footnote{More formally $\Delta T$ is the standard deviation in temperature among star-forming clouds measured by~\citet{0004-637X-822-2-59}. This was computed by fitting a normal distribution to the $25^{\rm th}$ and $75^{\rm th}$ percentiles.}.
This suggests $\Delta z \approx 4$, so this approximation likely does mean that the effect should be visible as a small correction which becomes larger as we look to increasing redshift.

\section{Discussion}
\label{sec:conclusion}

Both theory and observation strongly suggest that gas temperatures in star-forming regions of many galaxies should be higher than in star-forming regions of the Milky Way.  We have shown that failure to account for the resultant bottom-light IMF leads to an overestimation of both SFR and stellar mass for these galaxies.  This result has broad implications for both our observational and theoretical understanding of galaxy evolution.
The most important effects on current observations of galaxy evolution are as follows.
\begin{itemize}
\item{Every $z \gtrsim 6$ photometric stellar mass and star formation rate is overestimated by current fitting techniques, because the CMB increases the temperature in star-forming regions.  This may also help to explain the possible tension between the inferred stellar masses of high-redshift galaxies and theoretical halo mass functions \citep{Steinhardt2016}. However, several other explanations in which the UV luminosity to halo mass ratio would evolve sharply have also been proposed \citep{Finkelstein2015,Trac2015,Mashian2016}}.
\item{The stellar mass and star-formation rate of nearly every $z > 1$ star-forming galaxy are overestimated by current fitting techniques, because cosmic ray heating increases the temperature in star-forming regions.  Because this effect varies with SFR and therefore also with stellar mass for galaxies on the star-forming main sequence, a correction for it also alters the shape of inferred mass functions.  Tentative evidence of this has been seen in the centre of the Milky Way, which has both a top-heavy IMF~\citep{2013ApJ...764..155L} and enhanced cosmic ray density~\citep{goto_2013}, though the latter occurs over a somewhat larger region than the former. Cosmic ray heating may also contribute to an explanation for the apparent difference in the shapes of high-redshift stellar mass and halo mass functions \citep{Leauthaud2010,Leauthaud2012,Gonzalez2013,Behroozi2015,Steinhardt2016,Davidzon2017}.}
\item{The star-forming main sequence is narrower than previously believed.  At fixed mass and redshift, SFR overestimation is larger in galaxies with higher SFRs.  Thus correcting for this effect reduces the spread of the star-forming main sequence.}
\item{Accounting for these effects properly is difficult with current template fitting codes.  There are several codes~\citep[e.g. those of][]{LePhare,FAST,fsps} which allow different choices of IMF.  However, we find here that the correct choice of IMF is linked to other fit parameters, including redshift and SFR, which are currently treated independently.  In addition, modelling the existing stellar population requires not merely a star formation history but also a proper treatment of the linked variation in the IMF over the course of that history.}
\end{itemize}

Although these observational and analytical challenges are difficult, these same effects provide several tantalizing possibilities for the development of new models of feedback in galaxy evolution.  In particular, there should be strong feedback between star formation and temperature-driven changes to the IMF in star-forming galaxies.  An increased star-formation rate produces more cosmic rays, resulting in higher-temperature gas and dust.  An increased temperature in star-forming regions produces a more bottom-light IMF.  This increase in the fraction of massive stars in turn increases the cosmic ray density generated at fixed star-formation rate.  However, it also decreases the ability of molecular clouds to collapse so reducing the overall star-formation rate.  

Depending upon the magnitude of these two effects, two different behaviours are possible.  If the reduction in SFR is sharper than the increase in cosmic ray production per unit SFR, it provides negative feedback, leading to an equilibrium between star formation rate, the IMF and gas temperature.  It may be possible to develop a model in which these effects lead to an explanation for the observed star-forming main sequence as an equilibrium solution.  If so, there is copious data available to test such a model.  It should also be noted that the natural time-scale for such feedback would be the average gap between star formation and cosmic ray production, which, depending upon the IMF, should be of order $1$ to $3$\,Gyr because this is the typical age of stars with sufficient mass to produce cosmic rays.  This is longer than the dynamical time-scale for a typical star-forming galaxy but similar to the feedback time-scale estimated from the scatter in redshift of the star-forming main sequence \citet{Steinhardt2014b}.

However, if the increase in cosmic ray production outweighs the reduction in SFR, it instead produces positive feedback.  This runaway process produces a galaxy with very high gas and dust temperatures, an IMF consisting only of massive stars with short lifetimes and rapid production of dust and metals.  Although the true star formation rate in such a galaxy would be relatively low, the abundance of massive stars and their high luminosity would indicate a very young stellar population with a very high star formation rate if analyzed with current techniques with a Milky Way IMF.  This solution could be a reasonable description of the origin of starburst galaxies, which have many of these properties \citep{Weedman1981}.  To make a specific model for either case requires a treatment of the various cooling mechanisms in order to fully describe the feedback between changes in cosmic ray production and temperature.  At present, such models are very poorly constrained by extragalactic observations, and even less so at high redshift.

We shall consider a broader set of models in the future.
However, we note here that one generic prediction is that if all parameters except the SFR are fixed in a galaxy, then at lower SFRs there would be an equilibrium solution and at sufficiently high SFRs, the galaxy would instead enter the runaway regime. 
A galaxy following this path would first grow along an equilibrium track, potentially one that would be well-fit by the observed star-forming main sequence.  Galaxies growing along the main sequence increase their SFR over time, and therefore such a galaxy would eventually hit the runaway regime and become a starburst galaxy.  Finally, the rapid heating of gas would prevent the formation of new stars and the galaxy would become quiescent.  Furthermore, these transitions would be entirely secular, being driven by the long-term evolution of the galaxy rather than by external triggers.  Empirical models with some of these properties have been described in recent years \citep{Conroy2009b,Peng2010,Steinhardt2014b,Toft2014,Peng2015}, so this is a promising theoretical avenue to explore. It is possible that the temperature dependence of the IMF may bring the star-forming main sequence under the same theoretical umbrella as the starburst and quiescent regimes and it is certainly the case that continuing to neglect this phenomenon will yield misleading results.

\section*{Acknowledgements}

The authors thank Max Pettini for productive conversations on star formation as well as Rob Izzard for suggested reading on chemical evolution.
ASJ thanks the UK Marshall commission for financial support.  CLS thanks the ERC Consolidator Grant funding scheme (project ConTExt, grant number No. 648179) and the Carlsberg Foundation for support.
CAT thanks Churchill College for his fellowship.




\bibliographystyle{mnras}
\bibliography{ref} 

\begin{thebibliography}{}
\makeatletter
\relax
\def\mn@urlcharsother{\let\do\@makeother \do\$\do\&\do\#\do\^\do\_\do\%\do\~}
\def\mn@doi{\begingroup\mn@urlcharsother \@ifnextchar [ {\mn@doi@}
  {\mn@doi@[]}}
\def\mn@doi@[#1]#2{\def\@tempa{#1}\ifx\@tempa\@empty \href
  {http://dx.doi.org/#2} {doi:#2}\else \href {http://dx.doi.org/#2} {#1}\fi
  \endgroup}
\def\mn@eprint#1#2{\mn@eprint@#1:#2::\@nil}
\def\mn@eprint@arXiv#1{\href {http://arxiv.org/abs/#1} {{\tt arXiv:#1}}}
\def\mn@eprint@dblp#1{\href {http://dblp.uni-trier.de/rec/bibtex/#1.xml}
  {dblp:#1}}
\def\mn@eprint@#1:#2:#3:#4\@nil{\def\@tempa {#1}\def\@tempb {#2}\def\@tempc
  {#3}\ifx \@tempc \@empty \let \@tempc \@tempb \let \@tempb \@tempa \fi \ifx
  \@tempb \@empty \def\@tempb {arXiv}\fi \@ifundefined
  {mn@eprint@\@tempb}{\@tempb:\@tempc}{\expandafter \expandafter \csname
  mn@eprint@\@tempb\endcsname \expandafter{\@tempc}}}

\bibitem[\protect\citeauthoryear{{Arnouts}, {Cristiani}, {Moscardini},
  {Matarrese}, {Lucchin}, {Fontana}  \& {Giallongo}}{{Arnouts}
  et~al.}{1999}]{LePhare}
{Arnouts} S.,  {Cristiani} S.,  {Moscardini} L.,  {Matarrese} S.,  {Lucchin}
  F.,  {Fontana} A.,   {Giallongo} E.,  1999, \mn@doi [\mnras]
  {10.1046/j.1365-8711.1999.02978.x}, 310, 540

\bibitem[\protect\citeauthoryear{Bailin, Stinson, Couchman, Harris, Wadsley  \&
  Shen}{Bailin et~al.}{2010}]{0004-637X-715-1-194}
Bailin J.,  Stinson G.,  Couchman H.,  Harris W.~E.,  Wadsley J.,   Shen S.,
  2010, The Astrophysical Journal, 715, 194

\bibitem[\protect\citeauthoryear{Bastian, Covey  \& Meyer}{Bastian
  et~al.}{2010}]{doi:10.1146/annurev-astro-082708-101642}
Bastian N.,  Covey K.~R.,   Meyer M.~R.,  2010, \mn@doi [Annual Review of
  Astronomy and Astrophysics] {10.1146/annurev-astro-082708-101642}, 48, 339

\bibitem[\protect\citeauthoryear{{Bate}}{{Bate}}{2009}]{2009MNRAS.392..590B}
{Bate} M.~R.,  2009, \mn@doi [\mnras] {10.1111/j.1365-2966.2008.14106.x}, \href
  {http://adsabs.harvard.edu/abs/2009MNRAS.392..590B} {392, 590}

\bibitem[\protect\citeauthoryear{{Bate} \& {Bonnell}}{{Bate} \&
  {Bonnell}}{2005}]{2005MNRAS.356.1201B}
{Bate} M.~R.,  {Bonnell} I.~A.,  2005, \mn@doi [\mnras]
  {10.1111/j.1365-2966.2004.08593.x}, \href
  {http://adsabs.harvard.edu/abs/2005MNRAS.356.1201B} {356, 1201}

\bibitem[\protect\citeauthoryear{{Behroozi} \& {Silk}}{{Behroozi} \&
  {Silk}}{2015}]{Behroozi2015}
{Behroozi} P.~S.,  {Silk} J.,  2015, \mn@doi [\apj]
  {10.1088/0004-637X/799/1/32}, \href
  {http://adsabs.harvard.edu/abs/2015ApJ...799...32B} {799, 32}

\bibitem[\protect\citeauthoryear{{B{\"o}hm-Vitense}}{{B{\"o}hm-Vitense}}{1992}]{1992isa..book.....B}
{B{\"o}hm-Vitense} E.,  1992, {Introduction to stellar astrophysics. Volume 3.
  Stellar structure and evolution.}

\bibitem[\protect\citeauthoryear{{Bonnell}, {Clarke}  \& {Bate}}{{Bonnell}
  et~al.}{2006}]{2006MNRAS.368.1296B}
{Bonnell} I.~A.,  {Clarke} C.~J.,   {Bate} M.~R.,  2006, \mn@doi [\mnras]
  {10.1111/j.1365-2966.2006.10214.x}, \href
  {http://adsabs.harvard.edu/abs/2006MNRAS.368.1296B} {368, 1296}

\bibitem[\protect\citeauthoryear{{Bonnell}, {Larson}  \& {Zinnecker}}{{Bonnell}
  et~al.}{2007}]{2007prpl.conf..149B}
{Bonnell} I.~A.,  {Larson} R.~B.,   {Zinnecker} H.,  2007, Protostars and
  Planets V, \href {http://adsabs.harvard.edu/abs/2007prpl.conf..149B} {pp
  149--164}

\bibitem[\protect\citeauthoryear{Bothwell et~al.,}{Bothwell
  et~al.}{2017}]{doi:10.1093/mnras/stw3270}
Bothwell M.~S.,  et~al., 2017, \mn@doi [Monthly Notices of the Royal
  Astronomical Society] {10.1093/mnras/stw3270}, 466, 2825

\bibitem[\protect\citeauthoryear{{Bouwens} et~al.,}{{Bouwens}
  et~al.}{2015}]{Bouwens2015}
{Bouwens} R.~J.,  et~al., 2015, \mn@doi [\apj] {10.1088/0004-637X/803/1/34},
  \href {http://adsabs.harvard.edu/abs/2015ApJ...803...34B} {803, 34}

\bibitem[\protect\citeauthoryear{{Bouwens} et~al.,}{{Bouwens}
  et~al.}{2016}]{Bouwens2016}
{Bouwens} R.~J.,  et~al., 2016, \mn@doi [\apj] {10.3847/0004-637X/830/2/67},
  \href {http://adsabs.harvard.edu/abs/2016ApJ...830...67B} {830, 67}

\bibitem[\protect\citeauthoryear{Cappellari et~al.,}{Cappellari
  et~al.}{2012}]{cite-key}
Cappellari M.,  et~al., 2012, Nature, 484, 485

\bibitem[\protect\citeauthoryear{{Clarke} \& {Bromm}}{{Clarke} \&
  {Bromm}}{2003}]{2003MNRAS.343.1224C}
{Clarke} C.~J.,  {Bromm} V.,  2003, \mn@doi [\mnras]
  {10.1046/j.1365-8711.2003.06765.x}, \href
  {http://adsabs.harvard.edu/abs/2003MNRAS.343.1224C} {343, 1224}

\bibitem[\protect\citeauthoryear{{Conroy} \& {Gunn}}{{Conroy} \&
  {Gunn}}{2010}]{2010ApJ...712..833C}
{Conroy} C.,  {Gunn} J.~E.,  2010, \mn@doi [\apj]
  {10.1088/0004-637X/712/2/833}, \href
  {http://adsabs.harvard.edu/abs/2010ApJ...712..833C} {712, 833}

\bibitem[\protect\citeauthoryear{{Conroy} \& {Wechsler}}{{Conroy} \&
  {Wechsler}}{2009}]{Conroy2009b}
{Conroy} C.,  {Wechsler} R.~H.,  2009, \mn@doi [\apj]
  {10.1088/0004-637X/696/1/620}, \href
  {http://adsabs.harvard.edu/abs/2009ApJ...696..620C} {696, 620}

\bibitem[\protect\citeauthoryear{{Conroy}, {Gunn}  \& {White}}{{Conroy}
  et~al.}{2009b}]{fsps}
{Conroy} C.,  {Gunn} J.~E.,   {White} M.,  2009b, \mn@doi [\apj]
  {10.1088/0004-637X/699/1/486}, 699, 486

\bibitem[\protect\citeauthoryear{{Conroy}, {Gunn}  \& {White}}{{Conroy}
  et~al.}{2009a}]{2009ApJ...699..486C}
{Conroy} C.,  {Gunn} J.~E.,   {White} M.,  2009a, \mn@doi [\apj]
  {10.1088/0004-637X/699/1/486}, \href
  {http://adsabs.harvard.edu/abs/2009ApJ...699..486C} {699, 486}

\bibitem[\protect\citeauthoryear{{Davidzon} et~al.,}{{Davidzon}
  et~al.}{2017}]{Davidzon2017}
{Davidzon} I.,  et~al., 2017, \mn@doi [\aap] {10.1051/0004-6361/201730419},
  \href {http://adsabs.harvard.edu/abs/2017A%26A...605A..70D} {605, A70}

\bibitem[\protect\citeauthoryear{{Dias}, {Coelho}, {Barbuy}, {Kerber}  \&
  {Idiart}}{{Dias} et~al.}{2010}]{Dias2010}
{Dias} B.,  {Coelho} P.,  {Barbuy} B.,  {Kerber} L.,   {Idiart} T.,  2010,
  \mn@doi [\aap] {10.1051/0004-6361/200912894}, \href
  {http://adsabs.harvard.edu/abs/2010A%26A...520A..85D} {520, A85}

\bibitem[\protect\citeauthoryear{{Doane} \& {Mathews}}{{Doane} \&
  {Mathews}}{1993}]{Doane1993}
{Doane} J.~S.,  {Mathews} W.~G.,  1993, \mn@doi [\apj] {10.1086/173509}, \href
  {http://adsabs.harvard.edu/abs/1993ApJ...419..573D} {419, 573}

\bibitem[\protect\citeauthoryear{{Eggleton}, {Fitchett}  \& {Tout}}{{Eggleton}
  et~al.}{1989}]{1989ApJ...347..998E}
{Eggleton} P.~P.,  {Fitchett} M.~J.,   {Tout} C.~A.,  1989, \mn@doi [\apj]
  {10.1086/168190}, \href {http://adsabs.harvard.edu/abs/1989ApJ...347..998E}
  {347, 998}

\bibitem[\protect\citeauthoryear{{Fagotto}, {Bressan}, {Bertelli}  \&
  {Chiosi}}{{Fagotto} et~al.}{1994}]{Fagotto1994}
{Fagotto} F.,  {Bressan} A.,  {Bertelli} G.,   {Chiosi} C.,  1994, \aaps, \href
  {http://adsabs.harvard.edu/abs/1994A%26AS..105...29F} {105}

\bibitem[\protect\citeauthoryear{{Finkelstein} et~al.,}{{Finkelstein}
  et~al.}{2015}]{Finkelstein2015}
{Finkelstein} S.~L.,  et~al., 2015, preprint, \href
  {http://adsabs.harvard.edu/abs/2015arXiv150400005F} {} (\mn@eprint {arXiv}
  {1504.00005})

\bibitem[\protect\citeauthoryear{Gardner et~al.,}{Gardner et~al.}{2006}]{JWST}
Gardner J.~P.,  et~al., 2006, \mn@doi [Space Science Reviews]
  {10.1007/s11214-006-8315-7}, 123, 485

\bibitem[\protect\citeauthoryear{{Gonzalez}, {Sivanandam}, {Zabludoff}  \&
  {Zaritsky}}{{Gonzalez} et~al.}{2013}]{Gonzalez2013}
{Gonzalez} A.~H.,  {Sivanandam} S.,  {Zabludoff} A.~I.,   {Zaritsky} D.,  2013,
  \mn@doi [\apj] {10.1088/0004-637X/778/1/14}, \href
  {http://adsabs.harvard.edu/abs/2013ApJ...778...14G} {778, 14}

\bibitem[\protect\citeauthoryear{Goto}{Goto}{2013}]{goto_2013}
Goto M.,  2013, \mn@doi [Proceedings of the International Astronomical Union]
  {10.1017/S1743921314001070}, 9, 429–433

\bibitem[\protect\citeauthoryear{{Grogin} et~al.,}{{Grogin}
  et~al.}{2011}]{CANDELS}
{Grogin} N.~A.,  et~al., 2011, \mn@doi [\apjs] {10.1088/0067-0049/197/2/35},
  \href {http://adsabs.harvard.edu/abs/2011ApJS..197...35G} {197, 35}

\bibitem[\protect\citeauthoryear{Guszejnov, Hopkins  \& Ma}{Guszejnov
  et~al.}{2017}]{Guszejnov17}
Guszejnov D.,  Hopkins P.~F.,   Ma X.,  2017, \mn@doi [Monthly Notices of the
  Royal Astronomical Society] {10.1093/mnras/stx2067}, 472, 2107

\bibitem[\protect\citeauthoryear{{Hartmann}}{{Hartmann}}{2002}]{2002ApJ...578..914H}
{Hartmann} L.,  2002, \mn@doi [\apj] {10.1086/342657}, \href
  {http://adsabs.harvard.edu/abs/2002ApJ...578..914H} {578, 914}

\bibitem[\protect\citeauthoryear{{Hernandez} \& {Ferrara}}{{Hernandez} \&
  {Ferrara}}{2001}]{2001MNRAS.324..484H}
{Hernandez} X.,  {Ferrara} A.,  2001, \mn@doi [\mnras]
  {10.1046/j.1365-8711.2001.04346.x}, \href
  {http://adsabs.harvard.edu/abs/2001MNRAS.324..484H} {324, 484}

\bibitem[\protect\citeauthoryear{{Hopkins}}{{Hopkins}}{2012}]{2012MNRAS.423.2037H}
{Hopkins} P.~F.,  2012, \mn@doi [\mnras] {10.1111/j.1365-2966.2012.20731.x},
  \href {http://adsabs.harvard.edu/abs/2012MNRAS.423.2037H} {423, 2037}

\bibitem[\protect\citeauthoryear{{Jappsen}, {Klessen}, {Larson}, {Li}  \& {Mac
  Low}}{{Jappsen} et~al.}{2005}]{2005A&A...435..611J}
{Jappsen} A.-K.,  {Klessen} R.~S.,  {Larson} R.~B.,  {Li} Y.,   {Mac Low}
  M.-M.,  2005, \mn@doi [\aap] {10.1051/0004-6361:20042178}, \href
  {http://adsabs.harvard.edu/abs/2005A%26A...435..611J} {435, 611}

\bibitem[\protect\citeauthoryear{Jeans}{Jeans}{1902}]{Jeans1}
Jeans J.~H.,  1902, \mn@doi [Philosophical Transactions of the Royal Society of
  London A: Mathematical, Physical and Engineering Sciences]
  {10.1098/rsta.1902.0012}, 199, 1

\bibitem[\protect\citeauthoryear{{Karakas} \& {Lattanzio}}{{Karakas} \&
  {Lattanzio}}{2014}]{2014PASA...31...30K}
{Karakas} A.~I.,  {Lattanzio} J.~C.,  2014, \mn@doi [\pasa]
  {10.1017/pasa.2014.21}, \href
  {http://adsabs.harvard.edu/abs/2014PASA...31...30K} {31, e030}

\bibitem[\protect\citeauthoryear{{Klessen}}{{Klessen}}{2001}]{2001ApJ...556..837K}
{Klessen} R.~S.,  2001, \mn@doi [\apj] {10.1086/321626}, \href
  {http://adsabs.harvard.edu/abs/2001ApJ...556..837K} {556, 837}

\bibitem[\protect\citeauthoryear{{Kriek}, {van Dokkum}, {Labb{\'e}}, {Franx},
  {Illingworth}, {Marchesini}  \& {Quadri}}{{Kriek} et~al.}{2009}]{FAST}
{Kriek} M.,  {van Dokkum} P.~G.,  {Labb{\'e}} I.,  {Franx} M.,  {Illingworth}
  G.~D.,  {Marchesini} D.,   {Quadri} R.~F.,  2009, \mn@doi [\apj]
  {10.1088/0004-637X/700/1/221}, 700, 221

\bibitem[\protect\citeauthoryear{{Kroupa}}{{Kroupa}}{2001}]{2001MNRAS.322..231K}
{Kroupa} P.,  2001, \mn@doi [\mnras] {10.1046/j.1365-8711.2001.04022.x}, \href
  {http://adsabs.harvard.edu/abs/2001MNRAS.322..231K} {322, 231}

\bibitem[\protect\citeauthoryear{{Krumholz}}{{Krumholz}}{2011}]{2011ApJ...743..110K}
{Krumholz} M.~R.,  2011, \mn@doi [\apj] {10.1088/0004-637X/743/2/110}, \href
  {http://adsabs.harvard.edu/abs/2011ApJ...743..110K} {743, 110}

\bibitem[\protect\citeauthoryear{{Laigle} et~al.,}{{Laigle}
  et~al.}{2016}]{Laigle2016}
{Laigle} C.,  et~al., 2016, \mn@doi [\apjs] {10.3847/0067-0049/224/2/24}, \href
  {http://adsabs.harvard.edu/abs/2016ApJS..224...24L} {224, 24}

\bibitem[\protect\citeauthoryear{{Larson}}{{Larson}}{1985}]{1985MNRAS.214..379L}
{Larson} R.~B.,  1985, \mn@doi [\mnras] {10.1093/mnras/214.3.379}, \href
  {http://adsabs.harvard.edu/abs/1985MNRAS.214..379L} {214, 379}

\bibitem[\protect\citeauthoryear{{Larson}}{{Larson}}{1998}]{1998MNRAS.301..569L}
{Larson} R.~B.,  1998, \mn@doi [\mnras] {10.1046/j.1365-8711.1998.02045.x},
  \href {http://adsabs.harvard.edu/abs/1998MNRAS.301..569L} {301, 569}

\bibitem[\protect\citeauthoryear{{Larson}}{{Larson}}{2005}]{2005MNRAS.359..211L}
{Larson} R.~B.,  2005, \mn@doi [\mnras] {10.1111/j.1365-2966.2005.08881.x},
  \href {http://adsabs.harvard.edu/abs/2005MNRAS.359..211L} {359, 211}

\bibitem[\protect\citeauthoryear{{Leauthaud} et~al.,}{{Leauthaud}
  et~al.}{2010}]{Leauthaud2010}
{Leauthaud} A.,  et~al., 2010, \mn@doi [\apj] {10.1088/0004-637X/709/1/97},
  \href {http://adsabs.harvard.edu/abs/2010ApJ...709...97L} {709, 97}

\bibitem[\protect\citeauthoryear{{Leauthaud} et~al.,}{{Leauthaud}
  et~al.}{2012}]{Leauthaud2012}
{Leauthaud} A.,  et~al., 2012, \mn@doi [\apj] {10.1088/0004-637X/744/2/159},
  \href {http://adsabs.harvard.edu/abs/2012ApJ...744..159L} {744, 159}

\bibitem[\protect\citeauthoryear{{Lee} \& {Hennebelle}}{{Lee} \&
  {Hennebelle}}{2018}]{2018A&A...611A..89L}
{Lee} Y.-N.,  {Hennebelle} P.,  2018, \mn@doi [\aap]
  {10.1051/0004-6361/201731523}, \href
  {http://adsabs.harvard.edu/abs/2018A%26A...611A..89L} {611, A89}

\bibitem[\protect\citeauthoryear{{Longmore} et~al.,}{{Longmore}
  et~al.}{2015}]{2015EAS....75...43L}
{Longmore} S.,  et~al., 2015, in Conditions and Impact of Star Formation.
  Edited by R. Simon, R. Schaaf and J. Stutzki. EAS Publications Series, Volume
  75-76, 2015, pp.43-48. pp 43--48, \mn@doi{10.1051/eas/1575006}

\bibitem[\protect\citeauthoryear{{Low} \& {Lynden-Bell}}{{Low} \&
  {Lynden-Bell}}{1976}]{1976MNRAS.176..367L}
{Low} C.,  {Lynden-Bell} D.,  1976, \mn@doi [\mnras] {10.1093/mnras/176.2.367},
  \href {http://adsabs.harvard.edu/abs/1976MNRAS.176..367L} {176, 367}

\bibitem[\protect\citeauthoryear{{Lu}, {Do}, {Ghez}, {Morris}, {Yelda}  \&
  {Matthews}}{{Lu} et~al.}{2013}]{2013ApJ...764..155L}
{Lu} J.~R.,  {Do} T.,  {Ghez} A.~M.,  {Morris} M.~R.,  {Yelda} S.,   {Matthews}
  K.,  2013, \mn@doi [\apj] {10.1088/0004-637X/764/2/155}, \href
  {http://adsabs.harvard.edu/abs/2013ApJ...764..155L} {764, 155}

\bibitem[\protect\citeauthoryear{{Lucatello}, {Gratton}, {Beers}  \&
  {Carretta}}{{Lucatello} et~al.}{2005}]{2005ApJ...625..833L}
{Lucatello} S.,  {Gratton} R.~G.,  {Beers} T.~C.,   {Carretta} E.,  2005,
  \mn@doi [\apj] {10.1086/428105}, \href
  {http://cdsads.u-strasbg.fr/abs/2005ApJ...625..833L} {625, 833}

\bibitem[\protect\citeauthoryear{{Magnelli} et~al.,}{{Magnelli}
  et~al.}{2014}]{Magnelli2014}
{Magnelli} B.,  et~al., 2014, \mn@doi [\aap] {10.1051/0004-6361/201322217},
  \href {http://adsabs.harvard.edu/abs/2014A%26A...561A..86M} {561, A86}

\bibitem[\protect\citeauthoryear{{Mashian}, {Oesch}  \& {Loeb}}{{Mashian}
  et~al.}{2016}]{Mashian2016}
{Mashian} N.,  {Oesch} P.~A.,   {Loeb} A.,  2016, \mn@doi [\mnras]
  {10.1093/mnras/stv2469}, \href
  {http://adsabs.harvard.edu/abs/2016MNRAS.455.2101M} {455, 2101}

\bibitem[\protect\citeauthoryear{{McGuire} et~al.,}{{McGuire}
  et~al.}{2016}]{2016ApJ...825..135M}
{McGuire} J.~T.~W.,  et~al., 2016, \mn@doi [\apj]
  {10.3847/0004-637X/825/2/135}, \href
  {http://adsabs.harvard.edu/abs/2016ApJ...825..135M} {825, 135}

\bibitem[\protect\citeauthoryear{Narayanan \& Dav\'e}{Narayanan \&
  Dav\'e}{2012}]{Desika2}
Narayanan D.,  Dav\'e R.,  2012, \mn@doi [Monthly Notices of the Royal
  Astronomical Society] {10.1111/j.1365-2966.2012.21159.x}, 423, 3601

\bibitem[\protect\citeauthoryear{Narayanan \& Dav\'e}{Narayanan \&
  Dav\'e}{2013}]{Desika1}
Narayanan D.,  Dav\'e R.,  2013, \mn@doi [Monthly Notices of the Royal
  Astronomical Society] {10.1093/mnras/stt1548}, 436, 2892

\bibitem[\protect\citeauthoryear{Oey}{Oey}{2011}]{Oey2011}
Oey M.~S.,  2011, \mn@doi [The Astrophysical Journal]
  {10.1088/2041-8205/739/2/l46}, 739, L46

\bibitem[\protect\citeauthoryear{{Offner}, {Clark}, {Hennebelle}, {Bastian},
  {Bate}, {Hopkins}, {Moraux}  \& {Whitworth}}{{Offner}
  et~al.}{2014}]{2014prpl.conf...53O}
{Offner} S.~S.~R.,  {Clark} P.~C.,  {Hennebelle} P.,  {Bastian} N.,  {Bate}
  M.~R.,  {Hopkins} P.~F.,  {Moraux} E.,   {Whitworth} A.~P.,  2014, \mn@doi
  [Protostars and Planets VI] {10.2458/azu_uapress_9780816531240-ch003}, \href
  {http://adsabs.harvard.edu/abs/2014prpl.conf...53O} {pp 53--75}

\bibitem[\protect\citeauthoryear{Papadopoulos}{Papadopoulos}{2010}]{PPP2010}
Papadopoulos P.~P.,  2010, ApJ, 720, 226

\bibitem[\protect\citeauthoryear{{Peng} et~al.,}{{Peng}
  et~al.}{2010}]{Peng2010}
{Peng} Y.-j.,  et~al., 2010, \mn@doi [\apj] {10.1088/0004-637X/721/1/193},
  \href {http://adsabs.harvard.edu/abs/2010ApJ...721..193P} {721, 193}

\bibitem[\protect\citeauthoryear{{Peng}, {Maiolino}  \& {Cochrane}}{{Peng}
  et~al.}{2015}]{Peng2015}
{Peng} Y.,  {Maiolino} R.,   {Cochrane} R.,  2015, \mn@doi [\nat]
  {10.1038/nature14439}, \href
  {http://adsabs.harvard.edu/abs/2015Natur.521..192P} {521, 192}

\bibitem[\protect\citeauthoryear{{Perley} et~al.,}{{Perley}
  et~al.}{2016}]{2016ApJ...817....8P}
{Perley} D.~A.,  et~al., 2016, \mn@doi [\apj] {10.3847/0004-637X/817/1/8},
  \href {http://adsabs.harvard.edu/abs/2016ApJ...817....8P} {817, 8}

\bibitem[\protect\citeauthoryear{{Pols}, {Izzard}, {Stancliffe}  \&
  {Glebbeek}}{{Pols} et~al.}{2012}]{2012A&A...547A..76P}
{Pols} O.~R.,  {Izzard} R.~G.,  {Stancliffe} R.~J.,   {Glebbeek} E.,  2012,
  \mn@doi [\aap] {10.1051/0004-6361/201219597}, \href
  {http://esoads.eso.org/abs/2012A%26A...547A..76P} {547, A76}

\bibitem[\protect\citeauthoryear{{Press} \& {Schechter}}{{Press} \&
  {Schechter}}{1974}]{Press1974}
{Press} W.~H.,  {Schechter} P.,  1974, \mn@doi [\apj] {10.1086/152650}, \href
  {http://adsabs.harvard.edu/abs/1974ApJ...187..425P} {187, 425}

\bibitem[\protect\citeauthoryear{Privon et~al.,}{Privon
  et~al.}{2017}]{0004-637X-835-2-213}
Privon G.~C.,  et~al., 2017, The Astrophysical Journal, 835, 213

\bibitem[\protect\citeauthoryear{Schnee, Li, Goodman  \& Sargent}{Schnee
  et~al.}{2008}]{0004-637X-684-2-1228}
Schnee S.,  Li J.,  Goodman A.~A.,   Sargent A.~I.,  2008, The Astrophysical
  Journal, 684, 1228

\bibitem[\protect\citeauthoryear{{Sheth}, {Mo}  \& {Tormen}}{{Sheth}
  et~al.}{2001}]{Sheth2001}
{Sheth} R.~K.,  {Mo} H.~J.,   {Tormen} G.,  2001, \mn@doi [\mnras]
  {10.1046/j.1365-8711.2001.04006.x}, \href
  {http://adsabs.harvard.edu/abs/2001MNRAS.323....1S} {323, 1}

\bibitem[\protect\citeauthoryear{Sliwa, Wilson, Aalto  \& Privon}{Sliwa
  et~al.}{2017}]{2041-8205-840-2-L11}
Sliwa K.,  Wilson C.~D.,  Aalto S.,   Privon G.~C.,  2017, The Astrophysical
  Journal Letters, 840, L11

\bibitem[\protect\citeauthoryear{{Somerville} \& {Dav{\'e}}}{{Somerville} \&
  {Dav{\'e}}}{2015}]{Somerville2015}
{Somerville} R.~S.,  {Dav{\'e}} R.,  2015, \mn@doi [\araa]
  {10.1146/annurev-astro-082812-140951}, \href
  {http://adsabs.harvard.edu/abs/2015ARA%26A..53...51S} {53, 51}

\bibitem[\protect\citeauthoryear{{Steinhardt} \& {Speagle}}{{Steinhardt} \&
  {Speagle}}{2014a}]{Steinhardt2014b}
{Steinhardt} C.~L.,  {Speagle} J.~S.,  2014a, \mn@doi [\apj]
  {10.1088/0004-637X/796/1/25}, \href
  {http://adsabs.harvard.edu/abs/2014ApJ...796...25S} {796, 25}

\bibitem[\protect\citeauthoryear{Steinhardt \& Speagle}{Steinhardt \&
  Speagle}{2014b}]{0004-637X-796-1-25}
Steinhardt C.~L.,  Speagle J.~S.,  2014b, The Astrophysical Journal, 796, 25

\bibitem[\protect\citeauthoryear{{Steinhardt} et~al.,}{{Steinhardt}
  et~al.}{2014}]{Steinhardt2014a}
{Steinhardt} C.~L.,  et~al., 2014, \mn@doi [\apjl]
  {10.1088/2041-8205/791/2/L25}, \href
  {http://adsabs.harvard.edu/abs/2014ApJ...791L..25S} {791, L25}

\bibitem[\protect\citeauthoryear{{Steinhardt}, {Capak}, {Masters}  \&
  {Speagle}}{{Steinhardt} et~al.}{2016}]{Steinhardt2016}
{Steinhardt} C.~L.,  {Capak} P.,  {Masters} D.,   {Speagle} J.~S.,  2016,
  \mn@doi [\apj] {10.3847/0004-637X/824/1/21}, \href
  {http://adsabs.harvard.edu/abs/2016ApJ...824...21S} {824, 21}

\bibitem[\protect\citeauthoryear{Svoboda et~al.,}{Svoboda
  et~al.}{2016}]{0004-637X-822-2-59}
Svoboda B.~E.,  et~al., 2016, The Astrophysical Journal, 822, 59

\bibitem[\protect\citeauthoryear{{Tanvir} et~al.,}{{Tanvir}
  et~al.}{2012}]{2012ApJ...754...46T}
{Tanvir} N.~R.,  et~al., 2012, \mn@doi [\apj] {10.1088/0004-637X/754/1/46},
  \href {http://adsabs.harvard.edu/abs/2012ApJ...754...46T} {754, 46}

\bibitem[\protect\citeauthoryear{{Timmes}, {Woosley}  \& {Weaver}}{{Timmes}
  et~al.}{1995}]{1995ApJS...98..617T}
{Timmes} F.~X.,  {Woosley} S.~E.,   {Weaver} T.~A.,  1995, \mn@doi [\apjs]
  {10.1086/192172}, \href {http://adsabs.harvard.edu/abs/1995ApJS...98..617T}
  {98, 617}

\bibitem[\protect\citeauthoryear{{Toft} et~al.,}{{Toft}
  et~al.}{2014}]{Toft2014}
{Toft} S.,  et~al., 2014, \mn@doi [\apj] {10.1088/0004-637X/782/2/68}, \href
  {http://adsabs.harvard.edu/abs/2014ApJ...782...68T} {782, 68}

\bibitem[\protect\citeauthoryear{{Tout}, {Pols}, {Eggleton}  \& {Han}}{{Tout}
  et~al.}{1996}]{1996MNRAS.281..257T}
{Tout} C.~A.,  {Pols} O.~R.,  {Eggleton} P.~P.,   {Han} Z.,  1996, \mn@doi
  [\mnras] {10.1093/mnras/281.1.257}, \href
  {http://adsabs.harvard.edu/abs/1996MNRAS.281..257T} {281, 257}

\bibitem[\protect\citeauthoryear{{Trac}, {Cen}  \& {Mansfield}}{{Trac}
  et~al.}{2015}]{Trac2015}
{Trac} H.,  {Cen} R.,   {Mansfield} P.,  2015, \mn@doi [\apj]
  {10.1088/0004-637X/813/1/54}, \href
  {http://adsabs.harvard.edu/abs/2015ApJ...813...54T} {813, 54}

\bibitem[\protect\citeauthoryear{{Vogelsberger} et~al.,}{{Vogelsberger}
  et~al.}{2014}]{Illustris}
{Vogelsberger} M.,  et~al., 2014, \mn@doi [\nat] {10.1038/nature13316}, \href
  {http://adsabs.harvard.edu/abs/2014Natur.509..177V} {509, 177}

\bibitem[\protect\citeauthoryear{{Volonteri}, {Reines}, {Atek}, {Stark}  \&
  {Trebitsch}}{{Volonteri} et~al.}{2017}]{2017arXiv170400753V}
{Volonteri} M.,  {Reines} A.,  {Atek} H.,  {Stark} D.~P.,   {Trebitsch} M.,
  2017, preprint, \href {http://adsabs.harvard.edu/abs/2017arXiv170400753V} {}
  (\mn@eprint {arXiv} {1704.00753})

\bibitem[\protect\citeauthoryear{{Weedman}, {Feldman}, {Balzano}, {Ramsey},
  {Sramek}  \& {Wuu}}{{Weedman} et~al.}{1981}]{Weedman1981}
{Weedman} D.~W.,  {Feldman} F.~R.,  {Balzano} V.~A.,  {Ramsey} L.~W.,  {Sramek}
  R.~A.,   {Wuu} C.-C.,  1981, \mn@doi [\apj] {10.1086/159133}, \href
  {http://adsabs.harvard.edu/abs/1981ApJ...248..105W} {248, 105}

\makeatother
\end{thebibliography}



\appendix


\bsp	
\label{lastpage}
\end{document}